\documentclass[twocolumn,twoside,slac_two]{revtex4}
\usepackage{graphicx}
\usepackage{fancyhdr}
\newcommand{\BR}{B\kern -0.1em R}
\newcommand{\phif}{$ \phi$~{\em factory\/} }
\newcommand{\Dafne}{DA\char8NE}

\newcommand{\CP}{{\rm CP\/}}
\newcommand{\CPT}{{\rm CPT\/}}

\newcommand{\ns}{{\,\rm ns}}

\newcommand{\fb}  {\ensuremath{{\rm fb^{-1}}}}
\newcommand{\bra}[1]{\ensuremath{\langle\,#1\,|}}
\newcommand{\ket}[1]{\ensuremath{|\,#1\,\rangle}}

\newcommand{\eV}{{e\kern-.07em V}}

\newcommand{\MeV}{{\rm \,M\eV}}
\newcommand{\GeV}{{\rm \,G\eV}}
\newcommand{\epsi}        {\ensuremath{\epsilon}}
\newcommand{\epsil}{\epsilon'}
\newcommand{\Reepe}{\ensuremath{{\Re \left(\epsil/\epsi\right)}}}

\newcommand{\pai}[1]{\ensuremath{\pi^{{#1}}}}
\newcommand{\kao}[1]{\ensuremath{K^{{#1}}}}
\newcommand{\ks}{\ensuremath{K_S}}
\newcommand{\kl}{\ensuremath{K_L}}
\newcommand{\kls}{\ensuremath{K_{L,S}}}
\newcommand{\ko}{\ensuremath{K^0}}
\newcommand{\Vusfo}   {\ensuremath{|V_{\rm us} f_{\rm +}(0)|}}

\newcommand{\Vus}   {\ensuremath{V_{\rm us}}}
\newcommand{\Vud}   {\ensuremath{V_{\rm ud}}}
\newcommand{\kmudue}[1]{\ensuremath{K^{{#1}}_{\mu2}}}

\newcommand{\ktau}[1] {\ensuremath{K^{{#1}}_{\tau }}}
\newcommand{\ktaup}[1] {\ensuremath{K^{{#1}}_{\tau '}}}

\newcommand{\kltre}[1] {\ensuremath{K^{{#1}}_{{\rm l}3}}}
\newcommand{\kldue}[1] {\ensuremath{K^{{#1}}_{{\rm l}2}}}
\newcommand{\kedue}[1] {\ensuremath{K^{{#1}}_{{\rm e}2}}}
\newcommand{\kptre}[1] {\ensuremath{K^{{#1}}_{\pi3}}}
\newcommand{\Dkedue}[1]{\ensuremath{K^{{#1}} \to e^{{#1}}\nu}}
\newcommand{\Dkmudue}[1]{\ensuremath{K^{{#1}} \to \mu^{{#1}}\nu}}
\newcommand{\Dkpidue}[1]{\ensuremath{K^{{#1}} \to \pi^{{#1}}\pi^{0}}}

\newcommand{\Dktau}[1] {\ensuremath{K^{{#1}} \to \pi^{#1}\pi^{+}\pi^{-}}}
\newcommand{\Dktaup}[1]{\ensuremath{K^{{#1}}\to\pi^{#1}\pi^0\pi^0}}
\newcommand{\DKSpippim}{\ensuremath{K_S\rightarrow\pi^+\pi^-}}
\newcommand{\Dkpnn}{\ensuremath{K \to \pi\nu\bar{\nu}}}
\newcommand{\Dkzeropnn}{\ensuremath{K_L \to \pi^0\nu\bar{\nu}}}
\newcommand{\Dkpmpnn}{\ensuremath{K^\pm \to \pi^\pm\nu\bar{\nu}}}

\begin{document}

\title{Kaon physics}

%

\author{B. Sciascia}
\affiliation{Laboratori Nazionali di Frascati - INFN, via E.Fermi 40, 00044 Frascati (Rome) Italy}
%

\begin{abstract}
At present, the main topics addressed by kaon physics are the unitarity test of CKM matrix via precision
measurements of the Cabibbo angle as well as precision tests of discrete symmetries: in particular,
study of possible CPT violations in a model-independent way through the Bell-Steinberger relation, or
through the measurement of charge asymmetries. Other interesting topics are related to the test of predictions
from chiral perturbation theory. Also status and prospects of the \Dkpmpnn\ and \Dkzeropnn\ decays
are discussed.

\end{abstract}

\maketitle

\thispagestyle{fancy}


\section{INTRODUCTION}

Last year has been full of new results in the kaon physics, which continues to contribute
to the development and the understanding of the Standard Model (SM) flavor structure.
Given the plenty of experimental results, only a part of these ones will be described in this review.
In the first two sections, the NA48/2 results on \kptre{\pm} decays will be presented: Section~\ref{Sec:CP_KPM} 
is devoted to the search for direct CP violation in \Dktau{\pm} decays (\ktau{\pm})~\cite{NA48/2:Ag_ktau}\cite{NA48/2:Ag_ktaup},
 while Section~\ref{Sec:CUSP} describes
the measurement of the \pai{}\pai{} scattering length extracted from the \pai{0}\pai{0} invariant mass
distribution of \Dktaup{\pm} decays (\ktaup{\pm})~\cite{CUSP:NA48meas}, 
a very important and 
 unexpected by product of the CP violation analysis.
The status of the determination of \Vus\ and the unitarity test of the CKM matrix is presented in Section~\ref{Sec:Vus}.
Section~\ref{Sec:BS} presents the Bell-Steinberger relation~\cite{bs:paper} which from the requirement of the unitarity
offers the possibility to study the CPT invariance and to test the basic assumptions of the quantum field theories.
In particular, recent results from the KLOE experiment improve the determination of the \CP- and \CPT- violating parameters
$\Re(\epsilon)$ and $\Im(\delta)$~\cite{KLOE:bs}.
The role of \kldue{\pm} decays in probing the $\mu$-e universality emphasized by the recent NA48/2 improvement on the
measurement of $\Gamma(K^\pm \to e \nu_e)/\Gamma(K^\pm \to \mu \nu_\mu)$ ratio~\cite{NA48/2:kl2}, 
is described in Section~\ref{Sec:kl2}. 
Finally, in Section~\ref{Sec:kpnunubar} the progress toward the study of the very rare decay \Dkzeropnn\ and the prospects
to make a decisive measurement of \Dkpmpnn\ at the CERN SPS, are presented.

\section{Search for direct CP violation in \Dktau{\pm} decays}
\label{Sec:CP_KPM}

CP violation, more than 40 years after its discovery, still plays a central role
in present and future investigations of particle physics. For a long time there was no significant progress,
until NA48~\cite{CP:NA48} and KTeV~\cite{CP:KTeV} experiments demonstrated the existence
of direct CP violation with high significance. This was done by measuring a non-zero \Reepe\ parameter in the 
decays of neutral kaons into two pions. In 2001, the B-factory experiments Babar~\cite{CP:Babar1} and 
Belle~\cite{CP:Belle1} measured CP violation
in the system of neutral B mesons and in 2004 also the direct CP violation in B decays has 
been found~\cite{CP:Babar2}\cite{CP:Belle2}.

In kaons, besides the \Reepe\ parameter measured in \kls\ into two pion decays, a promising complementary observable
is the asymmetry between \kao{+} and \kao{-} decays into three pions. Direct CP violation manifesting itself as an
asymmetry in two CP-conjugate decay amplitude (as in \kptre{\pm} case) is important as a strong qualitative test of
the way in which the SM accommodates CP violation.
Unfortunately, in general the predicted Dalitz-plot slope asymmetries are in the order of few units per million and
the quantitative effort to constrain the fundamental parameters of the theory is difficult due to 
non-perturbative hadronic effects. An intense theoretical program is under way to improve these predictions, allowing
the direct CP violation measurements to be used as strong quantitative constraints on the SM.

The \kptre{\pm} matrix element squared is usually parameterized by a polynomial expansion~\cite{PDG06}:
\begin{displaymath}
|M(u,v)|^2 \sim 1 + gu + hu^2 + kv^2 {\mbox ,}
\end{displaymath}
where $g$, $h$, and $k$ are the so called linear and quadratic Dalitz-plot slope parameters 
($|h|$,$|k| \ll |g|$) and the two Lorentz invariant kinematic variables $u$ and $v$ are defined as
\begin{eqnarray}
   u & = &\frac{s_3 - s_0}{m_\pi^2} {\mbox ,}\hspace{.4cm} v = \frac{s_2 - s_1}{m_\pi^2} {\mbox ,} \nonumber \\
   s_i & = & (P_k -P_i)^2  {\mbox ,}\hspace{.4cm} i = 1,2,3 {\mbox ;}\nonumber\\
   s_0 & = & \frac{s_1 + s_2 + s_3}{3}\nonumber{\mbox .}
\label{eq:k3p}
\end{eqnarray}
Here $m_\pi$ is the pion mass, $P_k$ and $P_i$ are the kaon and the pion four-momenta, the indexes $i$=1,2 correspond
to the two identical (``even'') pions and the index $i$ = 3 to the pion of different charge (the ``odd'' pion).
If CP is conserved, the decay amplitudes must be equal for \kao{+} and for \kao{-} separately. In case of CP violation,
the two different amplitudes could be studied through their asymmetry or, equivalently, by means of the asymmetry
of the linear parameter $g$:
\begin{equation}
A_g = (g^+ - g^-)/(g^+ + g^-)~\simeq~\Delta g/(2g)~{\mbox ,} 
\end{equation}
where $\Delta g$ is the slope difference and $g$ is the average slope.
The slope asymmetry is strongly enhanced with respect to the asymmetry of the integrated decay 
rate~\cite{CP3p:zero}.
The prediction for $A_g$ provided at first order by the Chiral Perturbation Theory and in 
the framework of the SM vary from a few 10$^{-6}$ to a 
few 10$^{-5}$~\cite{Handbook2}\cite{CP3p:uno}\cite{CP3p:due}.
Theoretical calculations involving processes beyond the SM allow a wider range for $A_g$ 
which can be as high as 10$^{-4}$~\cite{CP3p:tre}.

The NA48/2 experiment in the framework of kaon physics program at the CERN SPS, measured the 
asymmetries $A_g$ and $A^0_g$ in \Dktau{\pm} and \Dktaup{\pm} decays respectively.
With 1.67$\times$10$^9$ \ktau{\pm} decays, the difference in the linear slope parameter of
the Dalitz plot for \kao{+} and \kao{-} has been measured~\cite{NA48/2:Ag_ktau}:
$$
\Delta g = (-0.7\pm0.9_{stat}\pm0.6_{trg}\pm0.6_{syst})\times10^{-4}{\mbox .}
$$
Using for the Dalitz plot slope the PDG value $g$ = -0.2154 $\pm$ 0.0035~\cite{PDG06}, $\Delta g$
translates into the direct CP violation charge asymmetry:
\begin{eqnarray}
A_g & = & (1.7\pm2.1_{stat}\pm1.4_{trg}\pm1.4_{syst})\times10^{-4}\nonumber\\
    & = & (1.7\pm2.9)\times10^{-4} \nonumber
\end{eqnarray}

Regarding \Dktaup{\pm} decays, 4.7$\times$10$^7$ events have been used to measure the slope 
difference~\cite{NA48/2:Ag_ktau}:
$$
\Delta g^0 = (2.3\pm2.8_{stat}\pm1.3_{trg}\pm1.0_{syst}\pm0.03_{ext})\times10^{-4}{\mbox ;}
$$
the corresponding asymmetry is found to be
\begin{eqnarray}
A_g^0 & = & (1.8\pm2.2_{stat}\pm1.0_{trg}\pm0.8_{syst}\pm0.2_{ext})\times10^{-4}\nonumber\\
      & = & (1.8\pm2.6)\times10^{-4} \nonumber {\mbox .}
\end{eqnarray}
For both measurements the uncertainty due to the trigger is statistical in nature, and
the global precision obtained is limited mainly by the available statistics.
On the systematics side, a careful control of detector and beam line asymmetries has been done
to avoid such effects fake the charge asymmetry.
The results have more than one order of magnitude better precision than the previous
measurements, and are compatible with the SM predictions. These measurements have been
obtained using the 2003 data sample. Analysis of the 2004 data will double the sample and will
improve accordingly the uncertainties.

\section{Cusp-like effect in \Dktaup{\pm}}
\label{Sec:CUSP}

\begin{figure}
\centering
\includegraphics[width=85mm]{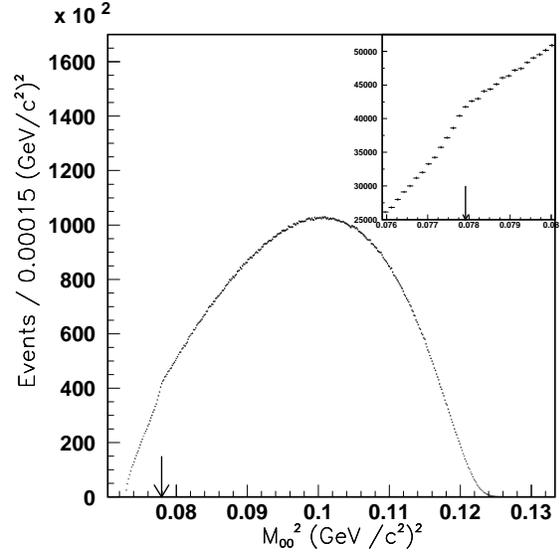}
\caption{$M_{00}^2$ distribution. The insert is an enlargement of a narrow region centered 
at $M_{00}^2 = (2m_+)^2$.} \label{fig:cusp}
\end{figure}

In searching for direct CP violation in \Dktaup{\pm}, NA48/2 experiment observed for a first
time a subtle and interesting phenomenon~\cite{CUSP:NA48meas}. In a partial sample of about 2.3$\times$10$^7$ 
\ktaup{\pm} decays acquired during 2003 data taking, the \pai{0}\pai{0} invariant mass distribution 
($M_{00}^2$) shows an anomaly in the region $M_{00}^2 = (2m_+)^2$ -$m_+$ is the \pai{\pm} mass- where a cusp-like
behavior is present (see figure~\ref{fig:cusp}).
This anomaly, observed thanks to the large statistical sample and the excellent $M_{00}^2$ resolution,
can be described by a re-scattering model~\cite{CUSP:Cabibbo} dominated by the contribution from the \Dktau{\pm} decay
through the charge-exchange reaction \pai{+}\pai{-} $\to$ \pai{0}\pai{0}.
The amplitude of the cusp is proportional to the $a_0 - a_2$ difference of the
\pai{}\pai{} S-wave scattering lengths. The quantity $a_0 - a_2$ is one of the few non-perturbative parameters which
can be predicted with excellent accuracy from first principles.
Recent calculations~\cite{CUSP:Isidori456} lead to a theoretical prediction of
$(a_0 - a_2)m_{\pi^+} = 0.265 \pm 0.004$ which has a precision not yet reached by the experiments.
The best direct information on \pai{}\pai{} scattering lengths is the one extracted from $K_{e4}$ decays by
the BNL-E865 experiment~\cite{CUSP:Isidori7}, which is affected by a statistical error of 6\%,
due to the intrinsic statistical limitation of the $K_{e4}$ decays with respect to the dominant \kptre{} modes;
moreover the BNL and NA48/2 determinations of $a_0 - a_2$ are affected by different theoretical and systematic errors. 
Fitting the $M_{00}$ distribution excluding re-scattering effects gives a 
$\chi^2$ = 13~574 for 148 degrees of freedom. A best fit to a re-scattering model
gives a good $\chi^2$ of 141 for 139 degrees of freedom, and provides a precise determination
of the $a_0 - a_2$ value: $(a_0 - a_2)m_{\pi^+} = 0.268 \pm 0.010_{stat} \pm 0.004_{syst}$, with
additional external uncertainty of $\pm 0.013_{ext}$ from branching ratio and theoretical 
uncertainties. If the correlation between $a_0$ and $a_2$ predicted by chiral symmetry is taken
into account, this result becomes 
$(a_0 - a_2)m_{\pi^+} = 0.264 \pm 0.006_{stat} \pm 0.004_{syst} \pm 0.013_{ext}$.

\begin{figure}
\centering
\includegraphics[width=8.5cm]{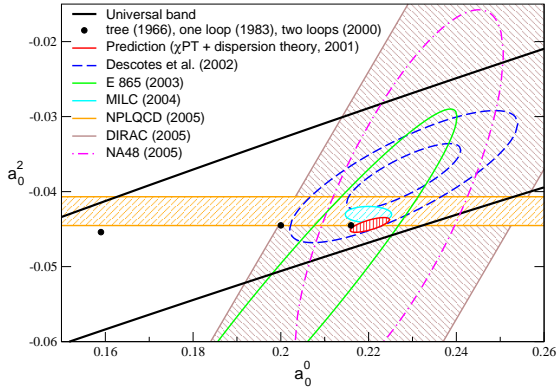}
\caption{Theoretical and experimental status of the S-wave scattering lengths.} \label{fig:cusp_colangelo}
\end{figure}

Figure~\ref{fig:cusp_colangelo} from~\cite{cusp:colangelo},
shows a summary of different theoretical determinations of the S-wave scattering lengths,
compared with present experimental measurements from NA48/2, CERN DIRAC and BNL E865;
the NA48/2 $a_0 - a_2$ determination has an uncertainty comparable with the two others available 
measurements.

\section{Unitarity test of CKM matrix and \Vus\ determination}
\label{Sec:Vus}

In the Standard Model, the quark weak charged current is
\begin{displaymath}
J_\alpha^+=(\bar u\ \bar c\ \bar t)\gamma_\alpha(1-\gamma_5)\,{\bf V}
\pmatrix{d\cr s\cr b\cr},
\end{displaymath} 
where {\bf V} is a $3\times3$ unitary matrix introduced by
Kobayashi and Maskawa \cite{KM73} in expansion on an original
suggestion by Cabibbo \cite{Cab63}.
The unitarity condition (${\bf V}^\dagger{\bf V}=1$) is required
by the assumption of universality of the weak interactions of leptons
and quarks and the absence of flavor-changing neutral currents.
The realization that a precise test of CKM unitarity can be obtained 
from the first-row constraint
$|V_{ud}|^2 + |V_{us}|^2 + |V_{ub}|^2 = 1$ (with $|V_{ub}|^2$ negligible)
has sparked a new interest in good measurements of quantities related to
$|V_{us}|$.
As we discuss in the following sections, $|V_{us}|$ can be determined 
using semileptonic kaon decays;
the experimental inputs are the Br's, lifetimes, 
and form-factor slopes. Both neutral (\ks\ or \kl) and charged kaons
may be used and provide independent measurements.
Many players have joined the game, as seen from Table~\ref{vusw}.\vglue2mm
\begin{table*}[t]
\begin{center}
\caption{Recent world data on $K_{\ell3}$ decays for calculation of $|V_{us}|$ }\vglue2mm
\begin{tabular}{@{}lll@{}}
\hline\hline
 Experiment & Measured quantities & References \\
\hline
 E865 & BR($\kao{+}\to\pai{0}_{\rm D}e^+\nu$)/BR($\kao{+}\to\pi^0_{\rm D}X^+$)~~~($\pai{0}_{\rm D}=\pai{0}\to e^+e^-\gamma$)
& \cite{E865+03:Ke3} \\
 KTeV & BR($K_{L\,e3}$), BR($K_{L\,\mu3}$), $\lambda_+(K_{L\,e3})$, $\lambda_{+,0}(K_{L\,\mu3})$ & \cite{KTeV+04:BR,KTeV+04:FF} \\
 ISTRA+ & $\lambda_+(K^-_{e3})$, $\lambda_{+,0}(K^-_{\mu3})$ & \cite{ISTRA+04:Ke3,ISTRA+04:Kmu3} \\
 NA48 & BR($K_{L\,e3}$)/BR(2 tracks), BR($K^\pm_{e3}$)/BR($\pi\pai{0}$), $\lambda_+(K_{L\,e3})$ & \cite{NA48+04:Ke3L,NA48+04:ICHEP,NA48+04:Ke3FF} \\
 KLOE & BR($K_{L\,e3}$), BR($K_{L\,\mu3}$), BR($K_{S\,e3}$), BR($K^\pm_{e3}$), BR($K^\pm_{\mu3}$), & 
        \cite{KLOE+06:KLBR,KLOE+06:KS2pi,KLOE+06:KSe3,eps} \\ 
      & $\lambda_+(K_{L\,e3})$, $\tau_L$, $\tau^\pm$ &  \cite{KLOE+06:KLe3FF,KLOE+05:KLlife} \\ 
\hline
\end{tabular}
\end{center}\label{vusw}
\end{table*}
KLOE is unique in that it is the only experiment that
can by itself measure the complete set of experimental inputs for the
calculation of $|V_{us}|$ using both charged and neutral kaons.
This is because the \phif\ is uniquely suited for measurements
of the \kl\ and \kao{\pm}\ lifetimes.
In addition, KLOE is the only experiment that can measure \ks\
Br's at the sub-percent level.

\subsubsection{Semileptonic Kaon Decays}
\label{sec:vusl3}
The semileptonic kaon decay rates
still provide the best means for the measurement of $|V_{us}|$
because only the vector part of the weak current contributes
to the matrix element $\bra{\pi}J_\alpha\ket{K}$. In general,
\begin{displaymath}
\bra{\pi}J_\alpha\ket{K} = f_+(t)(P+p)_\alpha + f_-(t)(P-p)_\alpha,
\end{displaymath}
where $P$ and $p$ are the kaon and pion four-momenta, respectively,
and $t=(P-p)^2$.
The form factors $f_+$ and $f_-$ appear because pions and kaons are 
not point-like particles, and also reflect both $SU(2)$ and $SU(3)$ 
breaking. For vector
transitions, the Ademollo-Gatto theorem \cite{ag} ensures that
$SU(3)$ breaking appears only to second order in $m_s-m_{u,d}$. 
In particular, $f_+(0)$ differs from unity by only 2--4\%. 
When the squared matrix element is evaluated, a factor of $m_\ell^2/m_K^2$
multiplies all terms containing $f_-(t)$, therefore 
can be neglected for $K_{e3}$ decays. For the description of $K_{\mu3}$ decays,
it is customary to use $f_+(t)$ and 
the scalar form factor $f_0(t) \equiv f_+(t) + [t/(m_K^2-m_\pi^2)]\,f_-(t)$. 

The semileptonic decay rates, fully inclusive of radiation, are given by
\begin{eqnarray}
\Gamma^i(K_{e3,\,\mu3}) & = & |V_{us}|^2\:{C_i^2\:G^2\:M^5\over768\pi^3}\:S_{\rm EW}\:I_{e3,\,\mu3} \label{eq:Gamsl}\\ 
                        &   & \times \:(1+\delta_{i,\,\rm em}+\delta_{i,\,SU(2)})\:|f^{\ko}_+(0)|^2.\nonumber
\end{eqnarray}
In the above expression, $i$ indexes $\ko\to\pi^\pm$ and $K^\pm\to\pai{0}$
transitions, for which $C_i^2 =1$ and 1/2, respectively. $G$ is the Fermi
constant, $M$ is the appropriate kaon mass, and $S_{\rm EW}$ is the
universal short-distance radiative correction factor \cite{as}. 
The $\delta$ terms are the long-distance
radiative corrections, which depend on the meson charges and lepton masses,
and the $SU(2)$-breaking corrections, which depend on the kaon charge 
\cite{aa}. 
The form factors are written as 
$f_{+,\,0}(t)=f_+(0)\tilde f_{+,\,0}(t)$, with $\tilde f_{+,\,0}(0)=1$. 
$f_+(0)$ reflects $SU(2)$- and $SU(3)$-breaking
corrections and is different for $\ko$ and $K^\pm$. $I_{e3,\,\mu3}$
is the integral of the Dalitz-plot density over the physical region
and includes $|\tilde f_{+,\,0}(t)|^2$.  $I_{e3,\,\mu3}$ does not 
account for photon emission; the effects of radiation are included
in the electromagnetic (em) corrections. The numerical factor
in the denominator of equation~\ref{eq:Gamsl}, $768=3\times2^8$, is chosen in
such a way that $I=1$ when the masses of all final-state particles vanish.
For $K_{e3}$, $I \approx 0.56$ and for $K_{\mu3}$, $I \approx 0.36$. 
The vector form factor $f_+$ is dominated by the vector $K\pi$ resonances, the
closest being the $K^*(892)$. Note that for $t>0$, $\tilde f_+(t)>1$. The
presence of the form factor increases the value of the phase-space integral
and the decay rate.
The natural form for $\tilde f_+(t)$ is
\begin{equation}
\tilde f_+(t) ={M_V^2\over M_V^2-t}.
\label{eq:pole}
\end{equation}
It is also customary to expand the form factor in powers of $t$ as
\begin{displaymath}
\tilde f_+(t)=1+\lambda'{t\over m^2_{\pi^+}}+
{\lambda''\over2}\left({t\over m^2_{\pi^+}}\right)^2.
\end{displaymath}
To compare the results obtained from each semileptonic decay mode
for both neutral and charged kaons without knowledge of $f_+(0)$, the relation~(\ref{eq:Gamsl})
is usually used to compute the quantity $f_+^{K^0}(0)\,|V_{us}|$.
This requires the $SU(2)$ and electromagnetic corrections for all four 
possible cases.

\subsection{ Results on semileptonic neutral kaon decays: KLOE, NA48, KTeV }
  Measurements of the absolute kaon branching ratios are a unique
 possibility of the \phif. At a \phif\  pairs of  monochromatic \ks
 \kl~ are produced  and the \kl (\ks) can be identified looking at the decay
 of 
the companion on the other side (tagging). The \kl~ is
 selected by identification of \ks~ decays while a \ks~
 beam is tagged using the fraction of events in which the \kl\ 
 interacts in the calorimeter (\kl~ crash).  Absolute branching ratios can be
 determined by counting the fraction of \kl's that decay into each
 channel and correcting for acceptances, reconstruction efficiencies,
 and background.  KLOE has measured the dominant \kl\ branching
 ratios using the \kl\ beam tagged by \DKSpippim\
 decays~\cite{KLOE+06:KLBR}.  In the 2001/2002 data sample, $\sim$13$\times
 10^6$ tagged \kl\ decays have been used for the measurement, and
 $\sim$4$\times 10^6$ to evaluate efficiencies.  To measure the BR's
 for decays to charged particles, a reconstructed \kl\ decay vertex is required
 in the DC fiducial volume. 
The number of events of each type is obtained by fitting the distribution
of missing momentum
minus missing energy, in the $\pi\mu$ mass
 assignment,
with the sum of the MC distributions   
for each of the decay channels.
To select \kl\ $\to$3\pai{0} events, at least three photons are required 
from the \kl\
decay vertex. The reconstruction efficiency and purity of the selected sample are both about 99\%.
The resulting BR's are: \\
$BR(K_{L}\to\pi e\nu(\gamma))=0.4049 \pm 0.0010 \pm 0.0031$,\\
$BR(K_{L} \to \pi \mu\nu(\gamma))=0.2726 \pm 0.0008 \pm 0.0022$,\\
$BR(K_{L} \to 3\pi^0)=0.2018 \pm 0.0004 \pm 0.0026$,\\
$BR(K_{L} \to \pi^{+}\pi^{-}\pi^{0}(\gamma))=0.1276 \pm 0.0006 \pm 0.0016$, \\
where the errors are dominated by the error on $\tau_L$ through 
the geometrical acceptance. 
Taking the BR's for rare \kl\ decays to $\pi^{+}\pi^{-}$, $\pi^{0}\pi^{0}$,
and $\gamma\gamma$ from the
PDG~\cite{PDG04} and imposing the constraint $\sum BR(\kl ) = 1$, 
$\tau_L$ can
be measured: $\tau_L = (50.72 \pm 0.17 \pm 0.33)$ \ns. 
Imposition of this constraint also results in more precise measurements 
of the dominant BR's:\\
$BR(K_{L} \to \pi e\nu(\gamma)) = 0.4007 \pm 0.0006 \pm 0.0014,$ \\
$BR(K_{L} \to \pi \mu\nu(\gamma)) = 0.2698 \pm 0.0006 \pm 0.0014,$ \\
$BR(K_{L} \to 3\pi^0) = 0.1997 \pm 0.0005 \pm 0.0019,$ \\
$BR(K_{L} \to \pi^{+}\pi^{-}\pi^{0}(\gamma))=0.1263 \pm 0.0005 \pm 0.0011$.
The \kl\ lifetime has been also measured directly~\cite{KLOE+05:KLlife}, 
employing 10$^7$ $K_{L} \to 3\pi^0$ events selected from the 2001/2002 data sample.
The result is $\tau_L = (50.92 \pm 0.17 \pm 0.25)$ \ns, which together with that from the
\kl\ BR measurements gives the KLOE average: $\tau_L = (50.84 \pm 0.23)$ \ns.

\begin{figure}
  \includegraphics[height=.3\textheight]{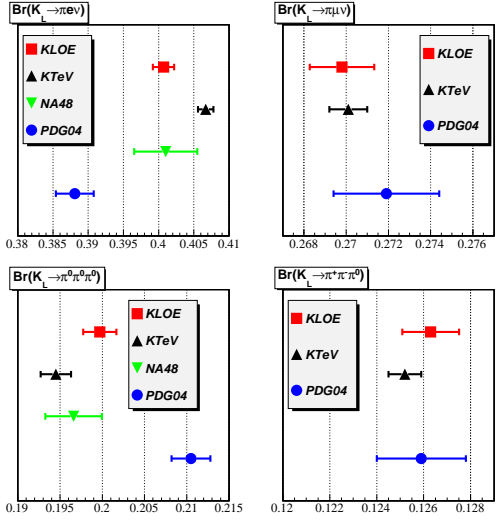}
  \caption{Branching ratios for the dominant \kl\ decays: comparison between different experimental measurements.}
\label{fig:anton}
\end{figure}

NA48 and KTeV experiments have used a different approach: they have
measured ratios of branching ratios since they do not have an absolute
normalization.  NA48 has normalized $K_{L} \to \pi e\nu$ events to
$K_{L} \to 2$ tracks. Using for BR($K_{L} \to 3\pi^0$) the average
of the PDG value and the KTeV measurement, they obtain $BR(K_{L} \to \pi
e\nu(\gamma)) = 0.401 \pm 0.004$ \cite{NA48+04:Ke3L}.  KTeV measures 5 ratios of  $K_{L}$ BR's
$\Gamma_{e3}$/$\Gamma_{\mu3}$, $\Gamma_{+-0}$/$\Gamma_{e3}$,
$\Gamma_{000}$/$\Gamma_{e3}$, $\Gamma_{+-}$/$\Gamma_{e3}$ and
$\Gamma_{00}$/$\Gamma_{000}$ . The 6 decay modes in the above ratios 
account for 99.93\%
of $K_{L}$ decays and the ratio can be combined to extract the
semileptonic branching ratio \cite{KTeV+04:BR}.
 In Fig.~\ref{fig:anton}
values of the branching ratios from
each experiment are reported and compared with values from  PDG.
 The new measurements agrees quite well with each other, while they disagree with the
old PDG values   for
the $K_{L} \to \pi e\nu$ and $K_{L} \to 3\pi^0$ decays.
\par
A measurement that is unique to KLOE is that of the BR for  $K_S$ semileptonic decays.
Using the 2001-2002 data set (410 pb$^{-1}$) a sample of 6\,500 events has been selected 
for each charge mode.
\begin{figure}[htb]
\centering
\includegraphics[height=6.5cm]{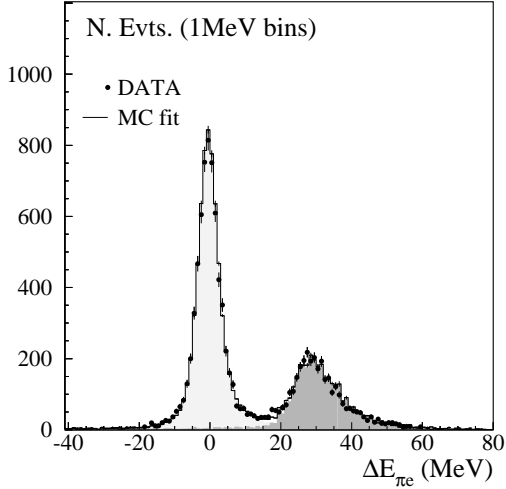}
\caption{$E_{miss}-P_{miss}$ distribution for $K_{S}\to\pi^{-}e^{+}\nu$ decays.}
\label{fig:kssemi}
\end{figure}
The distribution of the missing energy minus the missing momentum $\Delta E_{\pi e}$ is shown
in Fig.~\ref{fig:kssemi} for the $\pi^{-}e^{+}\nu$ sample. The signal is centered at zero
as expected for a missing neutrino. KLOE finds the value
$BR(K_{S}\to\pi e\nu)=(7.028\pm0.092)\times10^{-4}$~\cite{KLOE+06:KSe3}.
From the same sample also the charge asymmetry has been measured for the
first time. The charge asymmetry is found to be:
$$
A_{S}\!=\!  (1.5\pm9.6_{\mathrm{stat}}\pm 2.9_{\mathrm{syst}})\times 10^{-3}
$$ 
This result is compatible with that for $\kl$ semileptonic decays and with the
expectation obtained assuming $CPT$ symmetry, $A_{S}=2 Re(\epsilon)$.

\subsection{ Results on semileptonic charged kaon decays: KLOE, NA48, ISTRA+ }

In KLOE the measurement of the inclusive BR(K$^{\pm}_{l3}$)'s
uses four independent samples tagged by the decays:
$K_{\mu 2}^+$, $K_{\mu 2}^-$, $K_{\pi 2}^+$ and $K_{\mu 2}^-$.  
Two-body decays are rejected applying a cut on the charged decay particle
momentum computed in the kaon rest frame $p^{\ast}$.
The lepton squared mass, $m_{lept}^2$, is obtained from the velocity of 
the lepton computed from the ToF.
The number of $K_{e3}$ and $K_{\mu 3}$ decays is then obtained by
fitting the $m_{lept}^2$ distribution to a sum of MC distributions for
the signals and various background sources.
The BR is evaluated separately for each tag sample.
The resulting preliminary BR's are \cite{eps}:
\begin{center}
$BR(K^{\pm}_{e3})=
0.05047\pm 0.00046_{\rm stat}\pm0.00080_{\rm syst}$, \\
$BR(K^{\pm}_{\mu3}) =
0.03310\pm0.00040_{\rm stat}\pm0.00070_{\rm syst}$.
\end{center}
NA48 \cite{NA48/2:Kpme3} measures the $K^{+}_{e3}$ branching ratio normalizing to 
the $K_{\pi 2}$ decay taken from PDG04. 
They obtain: BR(${\rm K^{\pm}} \rightarrow \pi^o e \nu (\gamma)) = 
(5.14\pm0.02_{stat}\pm0.06_{syst})$\%. 
They also measures the $K^{+}_{e3} / K^{+}_{mu3}$ .

ISTRA+\cite{ISTRA+:Kpme3} has a similar approach, they obtain: 
BR(${\rm K^{\pm}} \rightarrow \pi^o e \nu (\gamma)) = (5.22\pm0.11)$\%.
It should be noticed here that  PDG06 value for the
$BR(K^{\pm} \rightarrow \pi^o  \pi^\pm)$ decreases by
$\approx$ 1 \%  due to the new $K_{e3}$ value from E865~\cite{E865+03:Ke3} 
and $K_{\mu2}$ from KLOE~\cite{KLOE+06:Kmu2}.

KLOE has also measured the ~\kao{\pm}~ lifetime. 
The kaon proper time is determined from the kaon momentum and path length,
which is evaluated taking into account the energy losses.
The vertex reconstruction efficiency and the 
resolution functions are measured directly on data by means
of the $\pi^o$ vertex reconstruction using only
calorimetric information.
The preliminary result is: $\tau (K^\pm) = 12.367 \pm 0.044_{Stat} \pm 0.065_{Syst}$ \ns\
in agreement with PDG~\cite{PDG06}.

\subsection {Form Factor}
The momentum dependence of the form factor, 
which is relevant for the integral over the phase space,
 is often described in terms of $\lambda$ slopes:
\\$f_+(t) = 1 + \lambda_+ t/m^2_{\pi^+}$ (linear) ,
\\$f_+(t) =
 1 + \lambda_+^{\prime} t/m^2_{\pi^+} +
 \frac{1}{2}\lambda_+^{\prime\prime}(t/m^2_{\pi^+})^2$
 (quadratic), 
\\or using the pole model: $f_+(t) = M_V^2/(M_V^2 - t)$.
The slopes and the pole are obtained fitting the t-spectrum with the corresponding expression
on form factor.

\begin{figure}
  \includegraphics[height=.30\textheight]{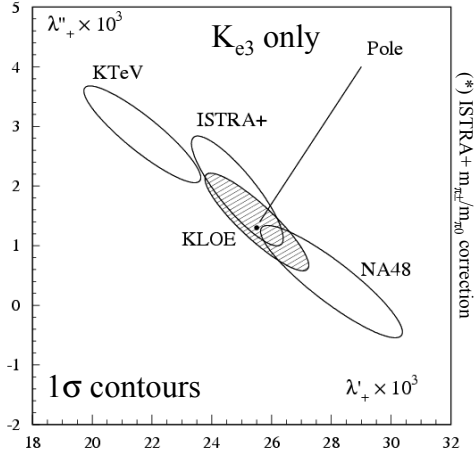}
  \caption{1$\sigma$ contours for different $\lambda^\prime_+$ and $\lambda^{\prime\prime}_+$ measurements.
           The cross gives the values obtained for a pole fit.}
\label{fig:anton1}
\end{figure}

KLOE experiment has fitted the $K_{e3}$ spectrum using the linear, the
 quadratic and the pole parametrization.
 A very slight preference for a small quadratic term is obtained
 \cite{KLOE+06:KLe3FF}. 
The results are:
\\{\bf quadratic} ~$\lambda_+^{\prime} = (25.5\pm 1.5 \pm 1.0)\times10^{-3}$ and
     $\lambda_+^{\prime\prime} = (1.4\pm 0.7 \pm 0.4)\times10^{-3}$ with $P(\chi^2)=92\% $,
\\{\bf pole model} ~$M_V = 870\pm 7$ \MeV\  with $P(\chi^2)=92.4\%$.
\\KTeV, ISTRA+ and NA48 experiments have fitted both $K_{e3}$ and $K_{\mu3}$ spectra.
KTeV experiment finds a quadratic slope term different from zero at 4 $\sigma$ level
and the slopes are consistent for $K_{e3}$ and $K_{\mu3}$ decay modes
\cite{KTeV+04:FF}. 
NA48 find no
evidence of quadratic term in the $K_{e3}$  spectrum, they obtain a $\lambda_0$
consistent with KTeV result in the $K_{\mu3}$ \cite{NA48+04:Ke3FF}. ISTRA+ experiment finds a 
 quadratic slope term different from zero at 2 $\sigma$ level and a
$ \lambda_0= (17.11\pm 2.31)\times10^{-3}$ \cite{ISTRA+04:Ke3,ISTRA+04:Kmu3}.
Fig.~\ref{fig:anton1}  compares the KLOE, KTeV, NA48, and ISTRA+ results. 
The values of the pole masses from the
KLOE, KTeV, and NA48 experiments are in agreement, and their
average value is $M_V = 875.3 \pm 5.4 ~~~~(\chi^2=1.8)$
to which correspond:
$\lambda_+^{\prime} = (25.42\pm 0.31)\times10^{-3} $ and
$\lambda_+^{\prime\prime} = (1.29\pm 0.03)\times10^{-3} $.

\subsection{V$_{us}$ extraction}

\begin{figure}
  \includegraphics[height=5.5cm]{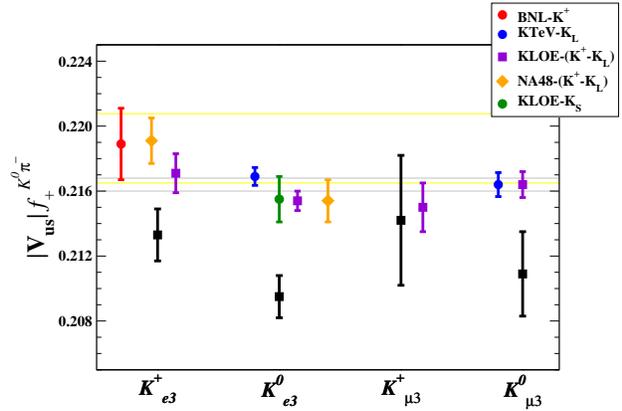}
  \caption{\Vusfo\ comparison between all measured semileptonic K decays. The black
           squares are the PDG 2004 values not included in the average.}
\label{fig:vusall}
\end{figure}

As already discussed, from equation~(\ref{eq:Gamsl}) \Vus\ can be determined using 
branching ratios, lifetimes and slopes as experimental inputs.
\Vus\ can be extracted using both charged and neutral modes, allowing for a consistency check between experiment 
and theory (see figure~\ref{fig:vusall}).
The value used for $\tau(K_L)$ is the
average between KLOE and PDG values. 
The slopes $\lambda_+^{\prime}$ and $\lambda_+^{\prime\prime}$
are those obtained by averaging
the pole masses from KTeV, NA48, and KLOE, while the slope $\lambda_o$
is the average from KTeV and ISTRA+. 
To extract \Vus\ the Leutwyler and Ross estimate of $f_+^{K^0 \pi^-}(0) = 0.961 \pm 0.008$~\cite{leutR}
has been used; this value has been confirmed recently by vastly improved latice QCD calculations~\cite{f0_Lattice}.

The grey band in figure~\ref{fig:vusall} is the average value for
\Vusfo\ = 0.2164(4), from which  ~\Vus = 0.2252(18) and it confirms
unitari\-ty within 1$\sigma$. The yellow band is obtained
imposing the unitarity using ~\Vud~=~0.97377(27)\cite{MS05}. The
uncertainty on $f_+^{K^o \pi^-}(0)$ is reflected in the
width of the ``uni\-ta\-rity'' band.

\subsubsection{$K\to\mu\nu$ Decays}
High-precision lattice quantum chromodynamics (QCD) results have 
recently become available and are
rapidly improving \cite{lat}. The availability of precise values for the
pion- and kaon-decay constants $f_\pi$ and $f_K$ allows to use of a relation
between $\Gamma(K_{\mu2})/\Gamma(\pi_{\mu2})$ and $|V_{us}|^2/|V_{ud}|^2$,
with the advantage that lattice-scale uncertainties and radiative corrections
largely cancel out in the ratio \cite{ref:marfk}:
\begin{equation}
{\Gamma(K_{\mu2(\gamma)})\over\Gamma(\pi_{\mu2(\gamma)})}=%
{|V_{us}|^2\over|V_{ud}|^2}\;{f_K^2\over f_\pi^2}\;%
{m_K\left(1-m^2_\mu/m^2_K\right)^2\over m_\pi\left(1-m^2_\mu/m^2_\pi\right)^2}\times C,
\label{eq:fkfp}
\end{equation}
where the precision of the numerical factor, C=(0.9930 $\pm$ 0.0035), due to structure-dependent 
corrections \cite{ref:fink} can be improved.
Thus, it could very well
be that the abundant decays of pions and kaons to $\mu\nu$ ultimately
give the most accurate determination of the ratio of $|V_{us}|$ to 
$|V_{ud}|$.
This ratio can be combined with direct measurements of $|V_{ud}|$ to obtain
$|V_{us}|$~\cite{ref:marfk,ref:fkfp}.
What is more interesting, however, is to combine all information
from $K_{e2}$, $K_{\mu2}$, $K_{e3}$, $K_{\mu3}$, and superallowed 
$0^+ \to 0^+$ nuclear $\beta$-decays to
experimentally test electron-muon and lepton-quark universality,
in addition to the unitarity of the quark mixing matrix. 

KLOE has measured  BR(K$_{\mu 2}$) using 175 $pb^{-1}$ collected in 2002, and
 found: $BR({\rm K^+} \rightarrow \mu^+\bar{\nu}(\gamma)) = 0.6366 \pm 0.0009_{ stat} \pm 0.0015_{ syst}$~\cite{KLOE+06:Kmu2}.
\begin{figure}
  \includegraphics[height=8.5cm]{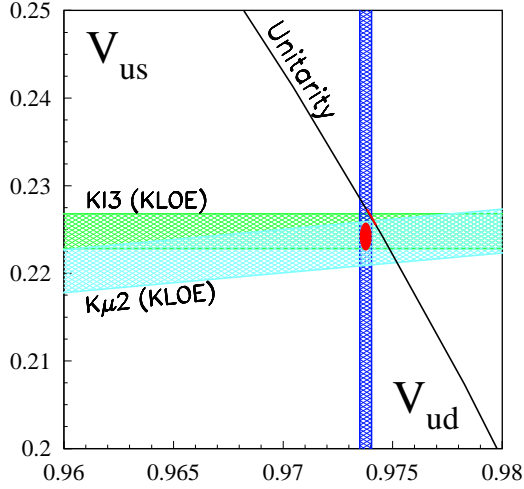}
  \caption{Comparison between $V_{us}/V_{ud}$ measurement from \Dkmudue{+} decays and
    \Vus~ from $K_{e3}$}
\label{fig:vusvud}
\end{figure}
From this measurement and using $f_K/f_{\pi}$ from MILC \cite{kmu2_theo}  collaboration
 we derive  $V_{us}/V_{ud} = 0.2294 \pm 0.0026$.
In fig.\ref{fig:vusvud} this result is compared with the one obtained from
 $K_{e3}$ decay and with the unitarity relation. From the fit, assuming unitarity, we obtain:
~\Vus = 0.2264(9).


\section{CP and CPT violation parameters using the Bell-Steinberger relation}
\label{Sec:BS}

The three discrete symmetries of quantum mechanics, charge conjugation ($C$), parity ($P$) and time reversal ($T$) are 
known to be violated in nature, both singly and in pairs. Only $CPT$ appears to be an exact symmetry of nature. 
Exact $CPT$ invariance holds in quantum field theory which assumes Lorentz invariance, 
locality and unitarity~\cite{CPT}. 
Testing the validity of $CPT$ invariance therefore probes the most fundamental assumptions of our present 
understanding of particles and their interactions.
These hypotheses are likely to be violated 
at very high energy scales, where quantum effects of the gravitational 
interaction cannot be ignored~\cite{CPTV}. 
On the other hand, since we still miss a consistent theory of quantum gravity,
it is hard to predict at which level violation of $CPT$ invariance might become experimentally observable. 

The neutral kaon system offers unique possibilities for the study of $CPT$ invariance. 
From the requirement of unitarity, Bell and Steinberger have derived a relation, the so called 
Bell-Steinberger relation (BSR)~\cite{bs:paper}. 
The most stringent limit on $CPT$-symmetry
violation is provided
by the mass difference 
$\left|m_{K^{0}}-m_{\bar{K}^{0}}\right|/m_{K^{0}}$.
This difference is related to the quantity
$\delta$ that parametrizes the $CPT$ violation in the mixing:
\begin{eqnarray}
  \nonumber
  \delta & = & 
  \frac{i\left(m_{K^{0}}-m_{\bar{K}^{0}}\right)
    +1/2\left(\Gamma_{K^{0}}-\Gamma_{\bar{K}^{0}}\right)}
	 {\Delta\Gamma} \times \\ 
	 & & \cos{\phi_{SW}} e^{i\phi_{SW}} 	 
	 \label{eq:defdelta}
\end{eqnarray}
where $\tan{\phi_{SW}}=2\Delta M/\Delta\Gamma$,
$\Delta\Gamma=\Gamma_{S}-\Gamma_{L}$, and 
$\Delta M=m_{L}-m_{S}$.
If we assume that the $CPT$ violation is negligible in the
decay ($\Gamma_{K^{0}}-\Gamma_{\bar{K}^{0}}=0$), we have:
\begin{equation}
  \nonumber
  \frac{m_{K^{0}}-m_{\bar{K}^{0}}}{m_{K^{0}}}=3\times10^{-14}\Im{\delta}.
\end{equation}
Violation at first order
in $m_{K}/M_{Planck}\sim 10^{-19}$, 
coming from quantum gravity effects,
can thus be tested by
putting an upper limit on $\Im{\delta}$ of order $\sim 10^{-5}$.

$\Re{\delta}$ and $\Im{\delta}$ are measured in two different ways.
$\Re{\delta}$ is measured both 
from the difference $A_{S}-A_{L}$ of the charge asymmetries
in $K_{S}$ and $K_{L}$ semileptonic decays,  
and from the time-dependent asymmetry
\begin{equation}
  \nonumber
  A_{CPT}=\frac{P(\bar{K}^{0}\to\bar{K}^{0}(t))-P(K^{0}\to K^{0}(t))}
  {P(\bar{K}^{0}\to\bar{K}^{0}(t))+P(K^{0}\to K^{0}(t))},
\end{equation}
where $P(K^{0}\to K^{0}(t))$ ($P(\bar{K}^{0}\to\bar{K}^{0}(t))$) 
is the probability for a $K^{0}$ ($\bar{K}^{0}$) 
to be observed as a $K^{0}$ ($\bar{K}^{0}$) after a time t.

$Im{\delta}$ is obtained from the unitarity relation (BSR):
\begin{eqnarray}
\nonumber
\left[\frac{\Gamma_{S}+\Gamma_{L}}{\Gamma_{S}-\Gamma_{L}}
  + i \tan{\phi_{SW}}\right]
\frac{\Re{\epsilon}-i\Im{\delta}}{1+\left|\epsilon\right|^{2}} &=& \\
 \frac{1}{\Gamma_{S}-\Gamma_{L}}\sum_{f}a^{*}_{S}(f)a_{L}(f) & = &
 \sum_{f}\alpha_{f}
 \label{eq:bellst}
\end{eqnarray}
where $a_{S,L}(f)$ are the $K_{S,L}$ decay amplitudes to the final
state $f$.

\subsection{Experimental inputs to $\Re{\delta}$}
\label{sec:redelta}

Using the time-dependent asymmetry in semileptonic decays~\cite{cplearpr} 
the CPLEAR Collaboration has obtained $\Re{\delta}$.
They measure the distribution
\begin{equation}
  A_{\delta}(t) = 4 \Re{\delta} + {\cal F}\left( \Im{\delta},y,\Re{x_{-}},\Im{x_{+}}\right) 
\end{equation}
that depends, beside the $CPT$-violating parameter $\delta$, on the 
parameters $y$ and $x_{\pm}$ that describe the $CPT$ violation in the decays
and the violation of the $\Delta S = \Delta Q$ rule, respectively.
They obtain:
\begin{eqnarray}
  \nonumber
  \Re{\delta} & = & (3.0\pm3.3\pm0.6)\times 10^{-4}
  \\
  \nonumber
  \Im{\delta} & = & (-1.5\pm2.3\pm0.3)\times 10^{-2}
  \\
  \nonumber
  \Re{x_{-}} & = & (0.2\pm1.3\pm0.3)\times 10^{-2}
  \\
  \Im{x_{+}} & = & (1.2\pm2.2\pm0.3)\times 10^{-2}  
\end{eqnarray}
This result is improved by adding as a constraint
the measurement~\cite{KLOE+06:KSe3,PDG04}
$A_{S}-A_{L}=4\left[\Re{\delta}+\Re{x_{-}}\right]=(-1.8\pm10.0)\times 10^{-3}$
to the original CPLEAR fit, obtaining:
\begin{eqnarray}
  \nonumber
  \Re{\delta} & = & (3.3\pm2.8)\times 10^{-4}
  \\
  \nonumber
  \Im{\delta} & = & (-1.1\pm0.7)\times 10^{-2}
  \\
  \nonumber
  \Re{x_{-}} & = & (-0.03\pm0.25)\times 10^{-2}
  \\
  \Im{x_{+}} & = & (0.8\pm0.7)\times 10^{-2}  
  \label{eq:imxpiu}
\end{eqnarray}
The uncertainty on $\Im{x_{+}}$ is reduced by a factor three.
As we will see in the next section, this parameter enters in the
unitarity relation and is one of the limiting factors in the determination
of $\Im{\delta}$ and $\Re{\epsilon}$.

\subsection{Experimental inputs to $\Im{\delta}$ and $\Re{\epsilon}$} 
\label{sec:imdelta}

The only decay modes relevant for the evaluation of
Eq.~(\ref{eq:bellst}) are the $\pi\pi(\gamma)$, $\pi\pi\pi$,
and semileptonic modes. The product of amplitudes $\alpha_{f}$
are obtained from the lifetimes of the kaons, the ratios
of $K_{S}$ and $K_{L}$ amplitudes $\eta$, and the branching ratios,
in the following way:
\begin{eqnarray}
\nonumber
\alpha_{\pi\pi}&=&\eta_{\pi\pi}BR(K_{S}\to\pi\pi)
\\
\nonumber
\alpha_{\pi\pi\gamma(DE)}&=&\sqrt{\frac{\tau_{S}}{\tau_{L}}BR(K_{L})BR(K_{S})}
e^{i\phi+-\gamma}
\\
\nonumber
\alpha_{\pi\pi\pi}&=&\frac{\tau_{S}}{\tau_{L}}\eta_{\pi\pi\pi}BR(K_{L}\to\pi\pi\pi)
\end{eqnarray} 
where $\pi\pi=\pi^{+}\pi^{-},\pi^{0}\pi^{0}$, 
and $\pi\pi\pi=\pi^{+}\pi^{-}\pi^{0},\pi^{0}\pi^{0}\pi^{0}$. The $\pi^{+}\pi^{-}$ mode
includes the inner bremsstrahlung (IB) emission. 
On the contrary, the contribution of the direct emission DE is estimated separately.

The contribution of semileptonic decays to Eq.~(\ref{eq:bellst})
is given by:
\begin{eqnarray}
  \nonumber
  \alpha_{\pi l\nu}&=&2\frac{\tau_{S}}{\tau_{L}}BR(K_{L}\to\pi l \nu)\times
  \\
  \nonumber
    & & \left[\Re{\epsilon}-\Re{y}-i\Im{\delta}-i\Im{x_{+}}\right]
  \\
  &=&\left[(A_{S}+A_{L})/4-i\Im{\delta}-i\Im{x_{+}}\right]
\end{eqnarray}
where $\Im{x_{+}}$ is given by Eq.~(\ref{eq:imxpiu}), and 
$(A_{S}+A_{L})/4=\Re{\epsilon}-\Re{y}=(-4.8\pm19.0)\times10^{-3}$,
while $\Im{\delta}$ is the unknown  of Eq.~(\ref{eq:bellst}).
The determination of $\Re{y}$ is of primary importance for the measurement of 
$A_{S}$~\cite{KLOE+06:KSe3}.

Some of the most recent measurements entering in the determination
of the $\alpha$ parameters are described in the following sections.
The numerical results are shown in Tab.~\ref{table:alphas}.
\begin{table*}[htb]
\caption{Values of the $\alpha$ parameters.}
\label{table:alphas}
\newcommand{\m}{\hphantom{$-$}}
\newcommand{\cc}[1]{\multicolumn{1}{c}{#1}}
\renewcommand{\tabcolsep}{2pc} 
\renewcommand{\arraystretch}{1.2} 
\begin{tabular}{@{}lll}
\hline
$\alpha$ parameters        &Real Part $\times 10^{3}$&Imaginary Part$\times 10^{3}$\\
\hline
$\alpha(\pi^{+}\pi^{-})$           &\m$1.126\pm0.014$&\m$1.064\pm0.014$\\
$\alpha(\pi^{0}\pi^{0})$           &\m$0.494\pm0.007$&\m$0.472\pm0.008$\\
$\alpha(\pi^{+}\pi^{-}\gamma(DE))$ &\m$0.000\pm0.002$&\m$0.000\pm0.002$\\
$\alpha(\pi l\nu)_{\Im{\delta}=0}$ &\m$0.003\pm0.002$&\m$-0.019\pm0.017$\\
$\alpha(\pi^{+}\pi^{-}\pi^{0})$    &\m$0.001\pm0.002$&\m$-0.001\pm0.002$\\
$\alpha(\pi^{0}\pi^{0}\pi^{0})$    &\m$<0.007$ at 95\% C.L.&\m$<0.007$ at 95\% C.L.\\
\hline
\end{tabular}\\[2pt]
\end{table*}

\subsubsection{$\pi\pi$}
The two body decays give the largest contribution to Eq.~(\ref{eq:bellst}).
The $K_{S}$ branching ratios are obtained from the recent
KLOE measurement of the ratio 
$\Gamma(K_{S}\to\pi^{+}\pi^{-})/\Gamma(K_{S}\to\pi^{0}\pi^{0})=(2.2549\pm0.0054)$~\cite{KLOE+06:KS2pi} 
with a sample of $400\times10^{6}$ $\phi\to K_{S}K_{L}$ decays.
The $K_{L}\to\pi^{+}\pi^{-}$ branching ratio has been recently measured by 
KLOE~\cite{klpipi} with $\sim45000$ events, they  find 
$BR(K_{L}\to\pi^{+}\pi^{-}(\gamma_{IB+DE}))=(1.963\pm0.021)\times10^{-3}$,
where both the IB and DE components are included.
The DE component is then subtracted using the KTeV result described
in the following.
KTeV~\cite{KTeV+04:BR} measured both the charged and neutral decays
with $\sim10^{5}$ events each. They find
$BR(K_{L}\to\pi^{+}\pi^{-}(\gamma_{IB}))=(1.975\pm0.012)\times10^{-3}$,
and
$BR(K_{L}\to\pi^{0}\pi^{0})=(0.865\pm0.010)\times10^{-3}$.
The measurement of charged mode includes only the IB component.
The two new measurements of $BR(K_{L}\to\pi^{+}\pi^{-})$ are in agreement
but disagree with the PDG value of 2004~\cite{PDG04}. 

\begin{figure}[htb]
\centering
\includegraphics[height=7.5cm]{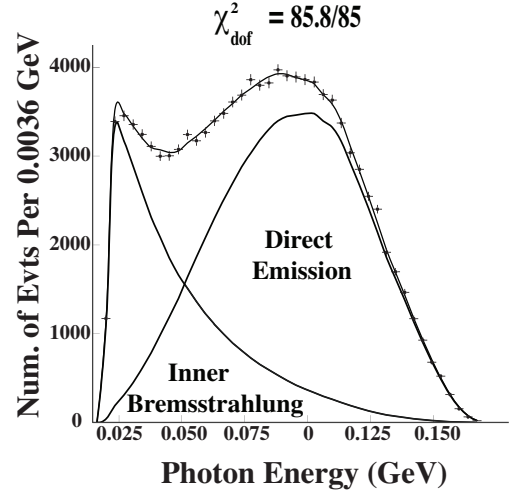}
\caption{Spectrum of photon energy in $K_{L}\to\pi^{+}\pi^{-}\gamma$ events,
measured by the KTeV Collaboration.}
\label{fig:ppgamma}
\end{figure}

The DE component is measured with two different methods. 
With the first method, the photon spectrum is measured directly
in $K_{L}\to\pi^{+}\pi^{-}\gamma$ decays. KTeV~\cite{ktevppgde} has 
measured it from a sample of $\sim10^{5}$ $\pi\pi\gamma$ events
with $E_{\gamma}^{*}>20$~\MeV\ (Fig.~\ref{fig:ppgamma}).
They find that the fraction of DE component for $E_{\gamma}^{*}>20$~\MeV\ 
is $DE/(DE+IB)=0.689\pm0.021$.
With the second method, the internal conversion of the photon into
a pair $e^{+}e^{-}$ is used. The resulting decay 
$K_{L}\to\pi^{+}\pi^{-}e^{+}e^{-}$ shows an asymmetry in the  
distribution of the angle between the $\pi^{+}\pi^{-}$ and 
$e^{+}e^{-}$ planes. This asymmetry is due to the interference between
the IB and DE components. KTeV~\cite{ktevppee} measures this
asymmetry with $\sim5000$ selected events and extracts
the information on the DE. The results are in agreement
with the previous method, but less precise.

The DE component in $K_{S}$ decay is suppressed by $CP$ violation. 
The $K_{S}$ photon spectrum has been measured in~\cite{ramberg}.

\subsubsection{$\pi l\nu$}

As has been described in Section~\ref{Sec:Vus}, 
the semileptonic decays of $K_{L}$ have been recently measured 
by KLOE~\cite{KLOE+06:KLBR}, KTeV~\cite{KTeV+04:BR}, and NA48~\cite{NA48+04:Ke3L}
in order to determine the Cabibbo angle. 
KLOE measured also $BR(K_{S}\to\pi e\nu)$ and the related charge asymmetry, $A_{S}$
(see Section~\ref{Sec:Vus}).

\subsubsection{$\pi\pi\pi$}

The amplitudes for $K\to3\pi$ decays can be decomposed
according to the $CP$ state of the three pion system:
\begin{eqnarray}
  \nonumber
  a_{L}&=&a_{L}^{CP-}(X,Y)
  \\
  a_{S}&=&a_{S}^{CP+}(X,Y)+a_{S}^{CP-}(X,Y)
  \label{eq:threepi}
\end{eqnarray}
where $a^{CP+(-)}$ is the amplitude for the
decay to the final state with $CP=+1(-1)$,
and $X\propto E_{+}-E_{-}$ and $Y\propto E_{0}$.
The decay to the $CP=+1$ state is suppressed
by the centrifugal barrier. Therefore, $a_{L}^{CP+}$ is
suppressed both by the centrifugal barrier and by $CP$ violation
and can be neglected. On the contrary, both $K_{S}$ amplitudes
are retained being one suppressed by $CP$ violation and the
other one by the centrifugal barrier.
The amplitudes have the following symmetry property:
\begin{equation}
  a^{CP\pm}(X,Y)=\mp a^{CP\pm}(-X,Y)
  \label{eq:symm3pi}
\end{equation}
Hence, the only contribution to Eq.~(\ref{eq:bellst}) 
comes from:
\begin{equation}
  \eta_{+-0}=\frac{\int dX dY a_{L}^{*}a_{S}^{CP-}}
      {\int dX dY |a_{L}|^{2}}
      \label{eq:eta}
\end{equation}
while from (\ref{eq:symm3pi}) follows:
\begin{equation}
  \nonumber
  \frac{\int dX dY a_{L}^{*}a_{S}^{CP+}}
      {\int dX dY |a_{L}|^{2}}=0
\end{equation}

CPLEAR measures $\eta_{+-0}$ from the time-dependent
asymmetry~\cite{cplearpr}:
\begin{equation}
  \nonumber
  \frac{\bar{N}_{3\pi}(t)-N_{3\pi}(t)}{\bar{N}_{3\pi}(t)+N_{3\pi}(t)}
\end{equation}
where $N_{3\pi}(t)$ ($\bar{N}_{3\pi}(t)$) is the number
of decays to three pions observed at time $t$ for 
kaons tagged as $K^{0}$ ($\bar{K}^{0}$) at time $t=0$.
They find:
\begin{equation}
  \nonumber
\eta_{+-0}=(-2\pm7)\times10^{-3}+i(-2\pm9)\times10^{-3}
\end{equation}

The $CP$-conserving component of the decay,
parametrized by the quantity: 
\begin{equation}
  \lambda=\frac{\int_{X>0} dX dY a_{L}^{*}a_{S}^{CP+}}
	       {\int_{X>0} dX dY |a_{L}|^{2}}
	       \label{eq:lambda}
\end{equation}
has been measured by NA48~\cite{na48ks3pi} by fitting the time-dependent
asymmetry:
\begin{equation}
  \nonumber
  \frac{N_{3\pi}^{X>0}(t)-N_{3\pi}^{X<0}(t)}
       {N_{3\pi}^{X>0}(t)+N_{3\pi}^{X<0}(t)}
\end{equation}
They find:
\begin{equation}
  \nonumber
\lambda=(0.038\pm0.010)+i(-0.013\pm0.007)
\end{equation}
A similar result has been obtained also by CPLEAR~\cite{cplearpr}.

\begin{figure}[htb]
  \centering
  \includegraphics[height=7.0cm]{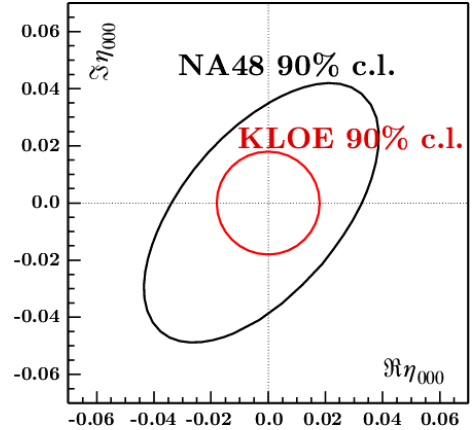}
  \caption{90\% C.L. limits on the plane 
    $\Im{\eta_{000}}$-$\Re{\eta_{000}}$ obtained
    from the measurements of NA48 and KLOE.}
  \label{fig:eta000}
\end{figure}
Because of its symmetry property $a_{S}^{CP+}(X,Y)=0$
for three equal pions, as in $K_{S}\to3\pi^{0}$ decays.
Therefore, this decay is $CP$ violating, and its
branching ratio is expected to be $\sim10^{-9}$.
Before the recent measurements of KLOE and NA48,
$\Im{\delta}$ was limited by the poor knowledge
of $\eta_{000}=a_{S}(000)/a_{L}(000)$.

There are two different ways of measuring this quantity.
NA48 measures the interference in $K\to3\pi^{0}$ decays as a function of the
proper time, with $5\times10^{6}$ decays from the 
{\it near target}, normalized to the rate of $10^{8}$ $K_{L}\to3\pi^{0}$
from the {\it far target}~\cite{na48ks3pi0}:
\begin{eqnarray}
  \nonumber
  f_{3\pi^{0}}(t)&\propto& 1+\left|\eta_{000}\right|^{2} 
  e^{-(\Gamma_{S}-\Gamma_{L})t}
  \\
  \nonumber
  & & + 2D(p)\left[\Re{\eta_{000}}\cos{\Delta m t} - \Im{\eta_{000}}\right.
\\
& & \left. \sin{\Delta m t} \right] e^{-1/2(\Gamma_{S}-\Gamma_{L})t}
\end{eqnarray}
where $D(p)$ is the {\it dilution} factor that describes the momentum dependent
production asymmetry between $K^{0}$ and $\bar{K}^{0}$ at the target.
NA48 measures:
\begin{equation}
  \eta_{000}=(-0.002\pm0.019)+i(-0.003\pm0.021)
\end{equation}

KLOE uses a pure $K_{S}$ beam tagged with a $K_{L}$.
This allows KLOE to directly search
$K_{S}\to3\pi^{0}$ decays~\cite{kloeks3pi0}.
The selection is done on a sample of $\sim5\times10^{8}$
$K_{L}-K_{S}$, looking for six photon clusters in the 
calorimeter and applying a veto for charged tracks in the
drift chamber. For each decay both $2\pi^{0}$ and
$3\pi^{0}$ hypotheses are tested defining two $\chi^{2}$-like variables.  
The events are counted in the signal box in the plane defined by
these two variables. KLOE finds 2 events with about 3 background events expected,
leading to the following 90\% limits on the branching ratio and
on the module of $\eta_{000}$:
\begin{eqnarray}
  \nonumber
  BR(K_{S}\to3\pi^{0}) &\leq& 1.2\times10^{-7}
  \\
  \left|\eta_{000}\right|&\leq& 0.018
\end{eqnarray}
The comparison of NA48 and KLOE results is shown in Fig.~\ref{fig:eta000}.

The $K_{L}$ decays to three pions have been measured
by KLOE and KTeV, and a preliminary result for the $3\pi^{0}$ decay
has been also presented by NA48. 

\subsection{Results}

\begin{figure*}[!t]
  \includegraphics[height=7.cm]{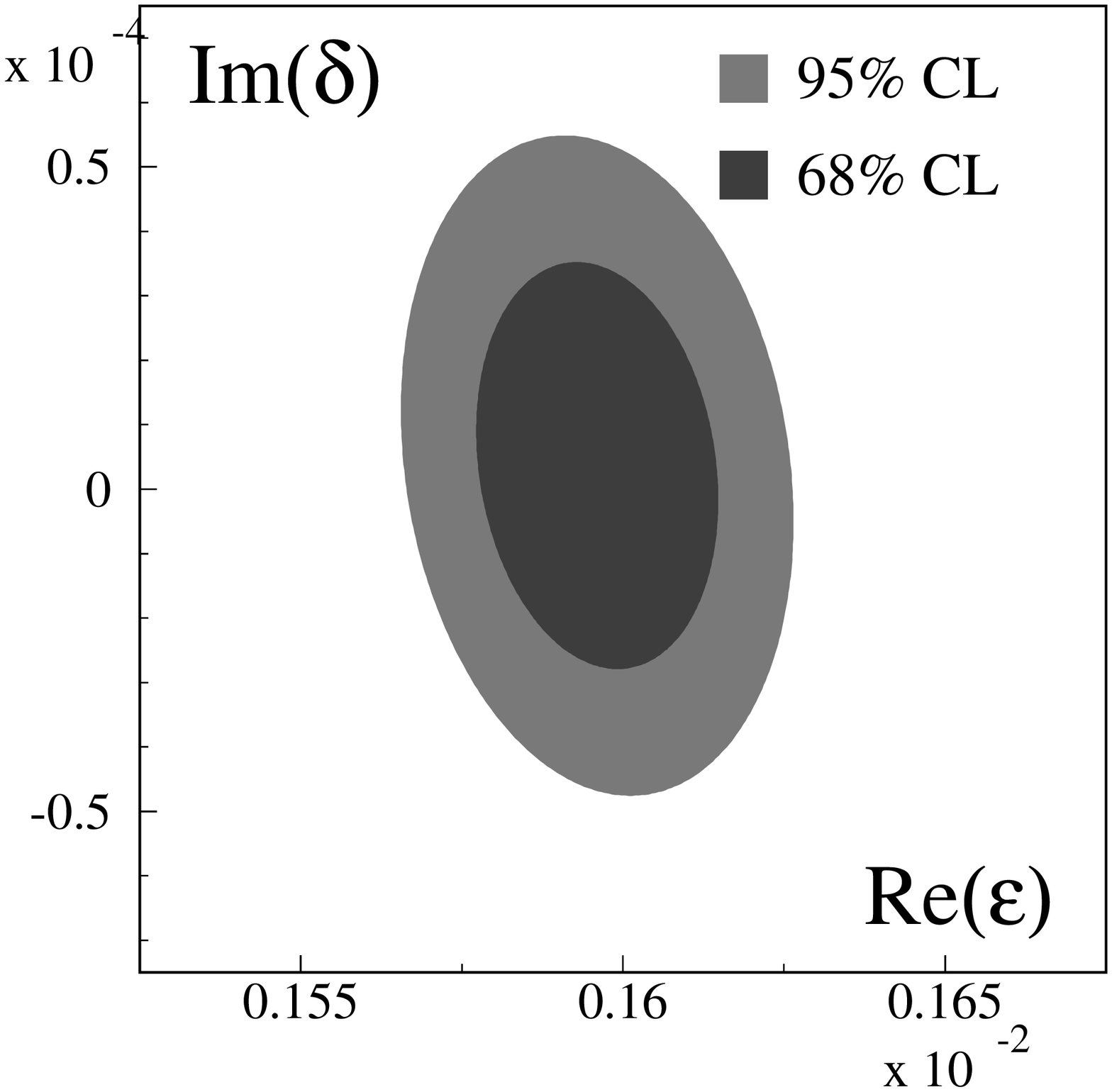}%
  \includegraphics[height=7.cm]{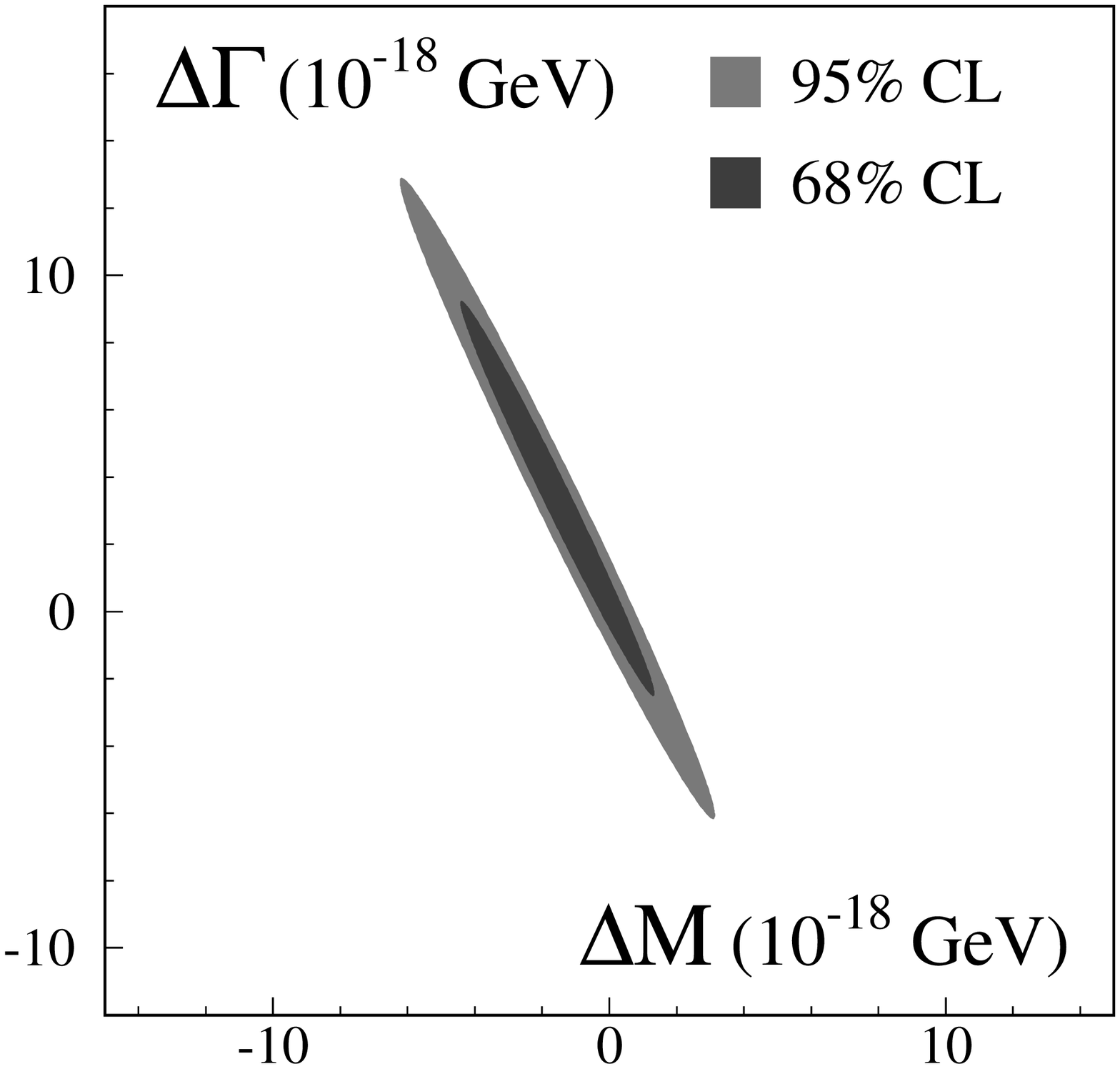}
  \caption{Allowed region at 68\% and 95\% C.L. in the $\Re{\epsilon}$,
    $Im{\delta}$ plane, and in the 
    $\Gamma_{K^{0}}-\Gamma_{\bar{K}^{0}}$,
    $m_{K^{0}}-m_{\bar{K}^{0}}$ plane.}
  \label{fig:deltavsespilon}
\end{figure*}

Following the method outlined in Sections~\ref{sec:redelta} and~\ref{sec:imdelta}
and the values in Tab.~\ref{table:alphas} 
the KLOE Collaboration together with G.~Isidori and G.~D'Ambrosio obtained
the following results:
\begin{eqnarray}
  \nonumber
  \Re{\epsilon} &=& (161.0\pm1.0)\times10^{-5}
  \\
  \Im{\delta} &=& (1.3\pm2.0)\times10^{-5}
  \label{eq:reepsimdelta}
\end{eqnarray} 
and from Eq.~(\ref{eq:defdelta}):
\begin{eqnarray}
  \nonumber
  \Gamma_{K^{0}}-\Gamma_{\bar{K}^{0}} &=& (5\pm4)\times10^{-18}~\mbox{GeV}
  \\
  m_{K^{0}}-m_{\bar{K}^{0}} &=& (-2.2\pm2.0)\times10^{-18}~\mbox{GeV}
  \label{eq:deltamdeltagamma}
\end{eqnarray}
The 90\% and 68\% C.L. regions are shown in Fig.~\ref{fig:deltavsespilon}.

Assuming no $CPT$ violation in the decay ($\Gamma_{K^{0}}-\Gamma_{\bar{K}^{0}}=0$), they find the 95\% C.L.:
\begin{equation}
  -4\times10^{-19} < m_{K^{0}}-m_{\bar{K}^{0}} < 7\times10^{-19}~\mbox{GeV}
  \label{eq:kloeresult}
\end{equation}
to be compared with the previous determination due to the
CPLEAR Collaboration:
\begin{equation}
  -10\times10^{-19} < m_{K^{0}}-m_{\bar{K}^{0}} < 17\times10^{-19}~\mbox{GeV}
  \label{eq:cplearresult}
\end{equation}

Eq.~(\ref{eq:kloeresult}) is at present the most stringent test of 
$CPT$ symmetry. 
The uncertainty is dominated by the knowledge of
the phases $\phi_{+-}$ and $\phi_{00}$, and of the
parameter $\Im{x_{+}}$. Further improvements are expected from
the analysis of the complete data sample of KLOE ($\sim2.5~\mbox{fb}^{-1}$),
in particular for what concerns the $K_{S}$ rare decays.    
However, there are no plans
to further improve the determination of $\phi_{+-}$, $\phi_{00}$.

\section{Test of lepton universality with \kldue{\pm} decays}
\label{Sec:kl2}

The measurement of the ratio R$_K$=\kedue{\pm}/\kmudue{\pm} between the branching ratios of \Dkedue{\pm} 
and \Dkmudue{\pm} decays is a sensible tool to test the lepton flavor universality and the V-A structure
of the weak interactions; similar arguments hold for pion physics with the ratio 
R$_\pi = \pi^\pm \to e^\pm \nu/\pi^\pm \to \mu^\pm \nu$ . 
Due to the uncertainties on non perturbative quantities like $f_K$,
the relevance of the single decays, like \kmudue{\pm}, in probing the SM is severely hindered.
In the ratio of the electronic and muonic decay modes, the hadronic uncertainties cancel to a very large extent, and
the SM prediction of R$_K$ is known with excellent accuracy~\cite{kl2:theo}.
This makes possible to exploit the good experimental precision on R$_K$ to constrain new physics effects.

\begin{figure}[htb]
  \includegraphics[height=6.5cm]{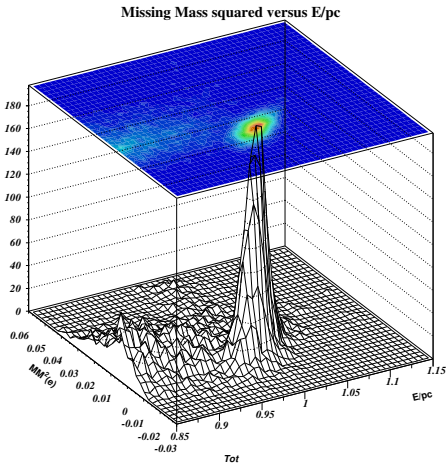}
  \caption{Missing mass squared versus E/p distribution of \kedue{\pm} candidates events, from NA48/2 experiment.}
  \label{fig:kl2}
\end{figure}

NA48/2 experiment at CERN SPS during 2003 data taking, has already collected more than 4 times the world 
total \kedue{\pm} statistics~\cite{NA48/2:kl2}. A similar amount of data has been collected in 2004.
The 2003 sample has been already analyzed, and 4\,670 \kedue{\pm} events have been found. 
In figure~\ref{fig:kl2} the missing mass squared versus E/p distribution for \kedue{\pm} candidates events is shown;
background contributions have been identified and subtracted.
The R$_K$ measurement is
\begin{equation}
\nonumber 
R_K = (2.416 \pm 0.043_{Stat} \pm 0.024_{Syst}) \times 10^{-5}{\mbox .}
\end{equation}
This value significantly improves the previous PDG value~\cite{PDG06} $R_K = (2.44 \pm 0.11) \times 10^{-5}$,
and will further improve with NA48/2 current analysis. Another $R_K$ measurement is in progress by the KLOE collaboration:
using the complete data set (2.5 \fb), 
which should have a statistical error comparable 
to the one of the NA48/2 and completely different contributions to the systematic error.

The NA48/2 measurement has to be compared with the SM prediction which reads
\begin{equation}
\nonumber 
R_K^{SM} = (2.416 \pm 0.043_{Stat} \pm 0.024_{Syst}) \times 10^{-5}{\mbox .}
\end{equation}
Denoting with $\Delta r_{NP}^{e-\mu}$ the deviation from $\mu-e$ universality in $R_K$ due to new physics, {\it i.e.}:
$R_K = R_K^{SM} \cdot (1 + \Delta r_{NP}^{e-\mu})$, the NA48/2 result requires at the 2$\sigma$ level,:
\begin{equation}
\nonumber 
-0.063 \le \Delta r_{NP}^{e-\mu} \le 0.017 {\mbox .}
\end{equation}
Two-body kaon decays are helicity suppressed in SM but generally unsuppressed in SM extensions, like the 
low-energy minimal SUSY extensions of the SM (MSSM) considered in~\cite{kl2:paradisi},
Here the question addressed is whether the SUSY can cause deviations from $\mu-e$ universality in \kldue{} decays
at a level which can be probed with the present attained experimental sensitivity (percent level).

\section{CKM unitarity and \Dkpnn\  decays}
\label{Sec:kpnunubar}

The \Dkpnn\ decays in the SM framework are treated in detail in a number of papers and reviews
(\textit{e.g} see ``Rare kaon decays'' in~\cite{PDG06} and references therein).
Here we recall some interesting aspects of these decays, before describing the status
of the measurement of the \Dkzeropnn\ and \Dkpmpnn\ branching ratios.

The unitarity of the CKM matrix assures the absence of Flavor Changing Neutral Current (FCNC) transitions 
at the tree level. FCNC processes can take place via loop diagrams containing internal quarks and intermediate bosons.
The semileptonic rare FCNC transitions \Dkzeropnn\ and \Dkpmpnn\ are particular and important because of
their clean theoretical character.
Firstly, the low energy hadronic matrix elements
required are just the matrix elements of quark currents between hadron states, which can be extracted with
good accuracy from non-rare semileptonic decays, \kltre{}. 
Moreover, the main contribution to the FCNC processes comes from the region of very small distances 
($\sim$ 1/$m_t$, 1/$m_Z$) where accurate estimate of strong interactions effects is possible
in the framework of perturbative QCD. Finally these processes are also sensitive to the contributions
from new heavy objects, \textit{e.g.} supersymmetric particles, therefore the comparison of the experimental
results with reliable theoretical estimates in the SM framework allows to search for new physics signals in 
these decays.
The \Dkzeropnn\ and \Dkpmpnn\ completely determine the apex the Unitarity Triangle~\cite{PDG06:ckm} and
the comparison with the Unitarity Triangle measured from B sector could provide decisive tests in the
flavor physics. Actually new physics may differentiate between the kaon and B-mesons sectors.
\begin{figure}[htb]
  \includegraphics[height=5.0cm]{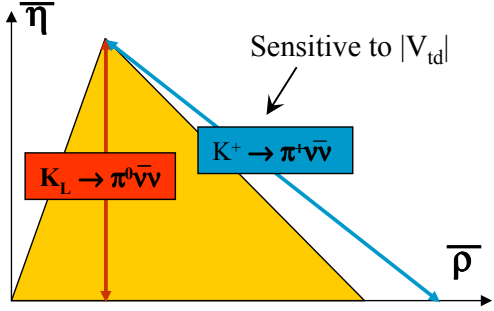}
  \caption{Determination of the Unitarity Triangle apex from \Dkpnn\ decays.}
  \label{fig:kpnn_triangle}
\end{figure}
At present the SM predictions of the two \Dkpnn{} decay rates are not extremely precise and read
\BR (\Dkpmpnn)$_{SM} = (8.8 \pm 1.1) \times 10^{-11}$~\cite{kpnn:brcarico} and 
\BR (\Dkzeropnn)$_{SM} = (3.0 \pm 0.6) \times 10^{-11}$~\cite{kpnn:brneutro}.

\begin{figure}[htb]
  \includegraphics[height=7.0cm]{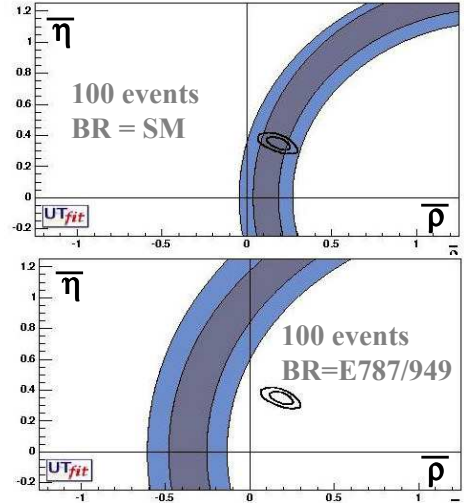}
  \caption{Ellipse: present determination of the unitarity triangle apex. Bands: ($\rho$,$\eta$) plane projection
of a 10~\% \BR (\Dkpmpnn) measurement with SM value (upper panel) or with BNL-787/949 value (lower panel).}
  \label{fig:kpnn_future}
\end{figure}

The two \Dkpmpnn\ candidate events observed by the BNL-787~\cite{kpnn:E787} experiment and
the one more candidate event observed by BNL-949~\cite{kpnn:E949} demonstrate that
the search for processes with branching ratios below $10^{-10}$ with missing energy although
very difficult is not impossible. The branching ratio inferred from these candidate events is
$(14.7^{+ 13.0}_{-8.9}) \times 10^{-11}$, with a central value higher than the SM prediction,
but compatible with this latter within the large errors. As sketched in figure~\ref{fig:kpnn_future},
a measurement of \BR (\Dkpmpnn) at the 10~\% level 
is required in order to match the theoretical uncertainty,
providing ground for precision tests of the SM flavor structure.
An indirect model-independent upper bound on \Dkzeropnn\ branching ratio can be set using the
\BR (\Dkpmpnn) value~\cite{kpnn:grossman_nir}: the BNL-787/949 measurement
gives a \BR (\Dkzeropnn) value which is about two order of magnitude greater than the SM expectation:
\begin{equation}\label{eq:grossman_nir}
\BR (\Dkzeropnn) \simeq 1.4  \times 10^{-9} \hspace{1.cm} (90~\%~C.L.)
\end{equation}
Recently the E391a experiment at the KEK proton synchrotron~\cite{kpnn:e391a} improved the experimental information on the
\Dkzeropnn\ decay setting a new upper limit of $2.1 \times 10^{-7}$ at 90~\% confidence level for the branching ratio, 
and this using only about 10~\% of the collected data in the first of the two data taking periods. This limit improves of a
factor 2.8 the previous limit~\cite{kpnn:kTeV} by kTeV at Fermilab which pioneered the so called \textit{pencil beam} technique.
The latter is the only method used nowadays to study the \Dkzeropnn\ decay and it consists in 
using a well collimated neutral kaon beam
in order to be able to use the beam direction as a kinematical constrain on the \kl\ momentum.
Analyzing the whole acquired statistics, the E391a experiment aims to reach the Grossman-Nir limit 
(see equation~(\ref{eq:grossman_nir})).
After the stopping of KOPIO experiment by US DoE, the future of \Dkzeropnn\ decay study relies on the step by step 
approach of J-PARC hadron facility. A Letter of Intent~\cite{kpnn:jparc_LoI} 
has been presented, which foresees firstly to move the E391a detector
at J-PARC, and afterwards the building of a new detector and a dedicated beam line with the aim to collect about 100 
\Dkzeropnn\ events, if SM holds

Because of the BNL-787/949 measurements, the situation is completely different for the \Dkpnn\ charged mode.
As already noted, a $\sim$100 events measurement could clarify the apparent discrepancy between the BNL measurement
and the SM theoretical predictions. Unfortunately the BNL program for the study of
\Dkpmpnn\ decay using stopped kaons was terminated prematurely. Five years ago at Fermilab, another initiative~\cite{kpnn:CKM},
called \textit{Charged Kaon at the Main injector} (CKM) and based on in-flight decay technique
to measure \Dkpmpnn\ with a statistics of 100 events, was proposed. The experiment was approved, but never funded
and eventually terminated.

\begin{figure}[htb]
  \includegraphics[height=5.5cm]{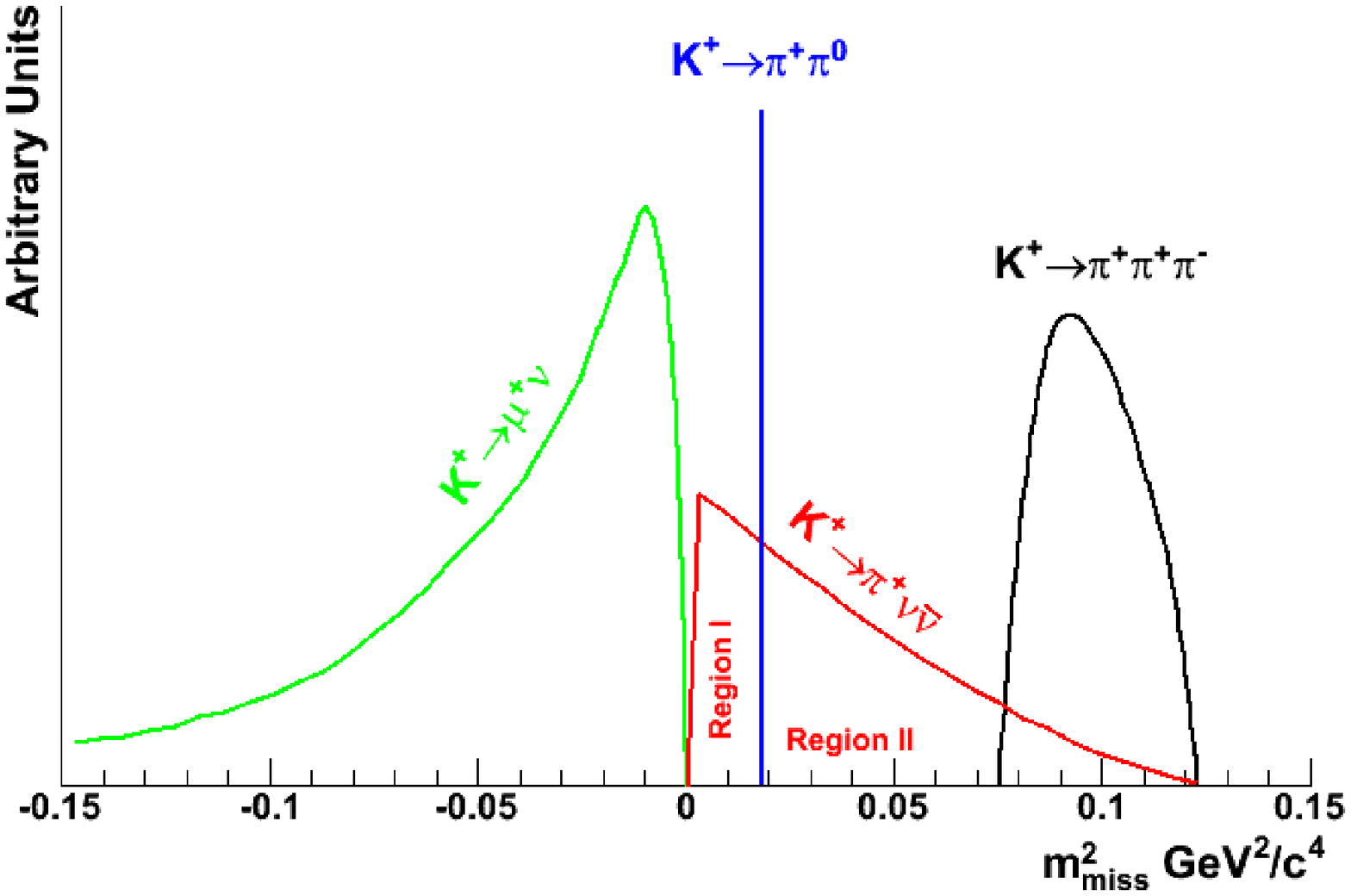}
  \caption{The missing mass spectrum for the \Dkpmpnn\ decay and the other main \kao{\pm} decays.}
  \label{fig:kpnn_missmass}
\end{figure}

The main background in studying \Dkpmpnn\ decays comes from the abundant \Dkpidue{\pm} events.
A kinematical constraint can be used (see regions I and II in figure~\ref{fig:kpnn_missmass}) to separate signal
from background. However events in which the \pai{0} is lost and the the \kao{\pm} and the \pai{\pm}
are not properly reconstructed can fake the signal. 
Moreover, due to the dependency of the photon detection efficiency from the kaon momentum, the use of a higher
kaon beam momentum can help in the \pai{0} rejection.
These considerations drove the proposal P326 (a.k.a. NA48/3)~\cite{kpnn:p326} for the CERN SPS which delivers protons of 400 \GeV.
The P326 strategy relies on a $10^{-8}$ \pai{0} rejection (this requires a single photon rejection of $10^{-5}$
for $E_\gamma > 1$~\GeV), on a redundant particle identification, and on the need of tracking and precisely timing
about 800~MHz of particles.
P326 aims to receive full approval by end of 2006 and
expects to collect about 100 events in two years with 10~\% of background.

\section{Conclusions}
There is plenty of new results in kaon physics since the 2005 Physics in Collision edition.
In this talk only a selection of these has been reported in.
The NA48/2 experiment searching for direct CP violation in charged kaon decays,
measured the asymmetries $A_g$ and $A_g^0$ respectively in \Dktau{\pm} and \Dktaup{\pm} decays.
With the present experimental accuracy, both asymmetries are compatible with zero.
The unexpected \pai{}\pai{} scattering length measurement in the \Dktaup{\pm} channel, 
opened a new way to test the structure of the \pai{}\pai{} interaction at low energies.
Using an up-to-date set of experimental measurements of form factors, branching ratios and lifetimes from
KLOE, KTeV, BNL E865, and NA48 experiments, a new determination of the Cabibo angle \Vus\ has been done
proving the unitarity of the CKM matrix at 1$\sigma$ level.
The NA48/2 preliminary measurement of $R_K$ and some theoretical works, renewed the interest in \kldue{\pm} decays
as particular interesting probe of new physics effects.
Finally the status and the perspectives for the golden-plated decays \Dkpnn\ have been presented, with the step by step
approach followed at KEK for the BR measurement of the \Dkzeropnn\ decay and the proposal to measure the \Dkpmpnn\ 
branching ratio at CERN SPS.

Kaons offer a unique playground to the test the Standard Model and to shed light on physics beyond the SM.

\bigskip 
\begin{acknowledgments}
I am grateful to G.~Isidori and F.~Mescia for illuminating discussions in preparing the talk presented at the conference.
I would like also to thank my colleagues of the KLOE collaboration A.~Antonelli, C.~Gatti, and M.~Moulson for 
their support in the drawing up of this manuscript.
\end{acknowledgments}

\bigskip 

\end{document}